\documentstyle[epsfig,12pt]{article}

\title{To the origin of the difference of FSI phases in $B\to\pi\pi$ and $B\to\rho\rho$ decays}
\author{A.B. Kaidalov\thanks{kaidalov@itep.ru} \  and
M.I. Vysotsky\thanks{vysotsky@itep.ru} \\ ITEP, Moscow, Russia}
\date{}
\begin{document}
\maketitle

\begin{abstract}
The final state interactions (FSI) model in which soft rescattering of low mass
intermediate states dominates is
suggested. It explains why the strong interaction phases  
are large in the
$B_d\to\pi\pi$ channel and are considerably smaller in the
$B_d\to\rho\rho$ one. Direct CP asymmetries of $B_d\to\pi\pi$
decays which are determined by FSI
phases are considered as well.
\end{abstract}

\section{Introduction}

There are three reasons to study FSI in $B$ decays: to
predict (or explain) the pattern of branching ratios, to
study strong interactions, and to forsee in what decays 
direct CPV will be large. In view of this necessity
a model for FSI in $B$ decays to two light mesons is suggested
and explored in the present paper.

The probabilities of three $B\to\pi\pi$ and three
$B\to\rho\rho$ decays are measured now with good accuracy. The
$C$-averaged branching ratios of these decays are presented in
Table 1 \cite{1}. Let us look at the ratio of the charge averaged
$B_d$ decay probabilities to the charged and neutral mesons:
\begin{equation}
R_\rho \equiv \frac{\rm Br(B_d
\to\rho^+\rho^-)}{\rm Br(B_d \to\rho^0\rho^0)} \approx
20 \; , \;\; R_\pi \equiv \frac{\rm Br(B_d
\to\pi^+\pi^-)}{\rm Br(B_d \to\pi^0\pi^0)} \approx 4
\;\; . \label{1}
\end{equation}

\begin{center}

{\bf Table 1}

\bigskip

\begin{tabular}{|c|c|c|c|}
\hline Mode & ${\rm Br}(10^{-6})$ &  Mode & ${\rm Br}(10^{-6})$ \\
\hline $B_d \to\pi^+ \pi^-$ & $5.2 \pm 0.2$ & $B_d \to
\rho^+\rho^-$ & $23.1 \pm 3.3$ \\ $B_d \to\pi^0 \pi^0$ & $1.3 \pm
0.2$ & $B_d \to \rho^0\rho^0$ & $1.16 \pm 0.46$ \\ $B_u \to\pi^+
\pi^0$ & $5.7 \pm 0.4$ & $B_u \to \rho^+\rho^0$ & $18.2 \pm 3.0$
\\ \hline
\end{tabular}

$C$-averaged branching ratios of $B\to\pi\pi$ and
$B\to\rho\rho$ decays.

\end{center}

\bigskip

The large difference of $R_\rho$ and $R_\pi$ is due to the
difference of FSI phases in $B\to\rho\rho$ and $B\to\pi\pi$
decays (see below). In Section 2 we will determine the 
differences of FSI
phases of tree amplitudes which describe $B\to\rho\rho$ and
$B\to\pi\pi$ decays into the states with isospins zero and two from
the data presented in Table 1. As a next step we will suggest a
mechanism which produces such phases. Once this
mechanism  is defined it becomes  possible to calculate FSI
phases of decay amplitudes into states with a definite isospin
(not only their differences).
A central question is: what intermediate states 
produce FSI phases in $B$-meson decays into two light mesons. 
In the weak decay $b\to u\bar u 
(d\bar d) d$ in the rest frame of a heavy quark (which is $B$-meson
rest frame as well) three fast light quarks are produced. 
Their energies are of the order of $M_B/3$ and momenta are 
more or less isotropically oriented. The energy of the fourth 
(spectator) quark is
of the order of $\Lambda_{QCD}$. This four quark state transforms
mainly
into multi pi-meson final state with the average pion multiplicity about
9 (this number follows from the experimentally known charged
particles multiplicity in $e^+e^-$ annihilation at $E_{cm}=3$GeV
multiplied by $1.5*1.5$ in order to take neutral pions and third quark jet
into account). The total branching ratio of such decays is about
$10^{-2}$. However such meson state does
not transform into the state composed from two light mesons moving into
opposite directions with momenta $M_B/2$. What meson state does
transform into two light mesons can be understood from the inverse 
reaction of two light meson scattering at the center of mass 
energy equal to the mass
of $B$-meson. The produced hadronic state consists of two jets of
particles moving in opposite directions. Each jet should originate 
from a quark-antiquark pair produced in the weak decay of $b$-quark. 
The square of invariant 
mass of a jet which contains spectator quark does not exceed  $M_B 
\Lambda_{QCD}$ and is much smaller than $M_B^2$. The energy of this jet
is determined by that of a companion quark and is about $M_B/2$. That
is why the square of invariant mass of the second jet also does not exceed 
$M_B \Lambda_{QCD}$. So for $B$-decays the mass of a hadron cluster 
which transforms into
light meson in the final state should not exceed 
1.5 GeV. Following these
arguments in the calculation of the imaginary
parts of the decay amplitudes we will take into account only two
(relatively light) particle intermediate states for which
branching ratios of $B$-meson are maximal.

In Section 3 we will calculate FSI phases of tree amplitudes
describing $B\to\pi\pi$ decays taking into account $\rho\rho$,
$\pi\pi$ and $\pi a_1$ intermediate states which by $t(u)$-channel
exchanges are converted into $\pi\pi$.
We will find that large probability of $B\to\rho^+\rho^-$ decay
explains about half of FSI phases of $B\to\pi\pi$ decays.
Relatively small probability of $B\to\pi^+\pi^-$ decay prevents
generation of noticeable FSI phase of $B \to\rho\rho$ amplitudes
through $B\to\pi^+\pi^- \to\rho\rho$ chain. 

We will demonstrate that the strong interaction phase of the penguin amplitude is opposite to the result of quark loop calculation, which is very important for the value of a
direct CPV asymmetry $C_{\pi^+\pi^-}\equiv C_{+-}$
discussed in Section 4.
Predictions for CPV asymmetries $C_{00}$ and  $S_{00}$ will be presented
in Section 4
as well and the value of the unitarity triangle angle $\alpha$ will
be extracted from the experimental data on CPV asymmetry $S_{+-}$.

Subject of rare $B$ decays is an object of intensive study nowadays
and an interested reader can find extensive list of references
in a recent paper \cite{7}.
\section{Phenomenology; $|\delta_0^\pi - \delta_2^\pi|$ and
$|\delta_0^\rho - \delta_2^\rho|$}

Let us present $B\to\pi\pi$ decay amplitudes in the so-called
``$t$-convention'', in which the penguin amplitude with the
intermediate $c$-quark multiplied by $V_{ub} V_{ud}^* + V_{cb} V_{cd}^*
+ V_{tb} V_{td}^* =0$ is subtracted from the decay amplitudes
\cite{2}:
\begin{eqnarray}
M_{\bar B_d \to \pi^+\pi^-} & = & \frac{G_F}{\sqrt 2}|V_{ub}
V_{ud}^*| m_B^2 f_\pi f_+(0) \left\{e^{-i\gamma}\frac{1}{2\sqrt
3}A_2 e^{i\delta_2^\pi} \right. + \nonumber \\ & + & e^{-i\gamma}
\left.\frac{1}{\sqrt 6} A_0 e^{i\delta_0^\pi} +
\left|\frac{V_{td}^* V_{tb}}{V_{ub} V_{ud}^*}\right| e^{i\beta} P
e^{i(\delta_P^\pi + \tilde\delta_0^\pi)}\right\} \;\; , \label{2}
\end{eqnarray}
\begin{eqnarray}
M_{\bar B_d \to \pi^0\pi^0} & = & \frac{G_F}{\sqrt 2}|V_{ub}
V_{ud}^*| m_B^2 f_\pi f_+(0) \left\{e^{-i\gamma}\frac{1}{\sqrt
3}A_2 e^{i\delta_2^\pi} \right. - \nonumber \\ & - & e^{-i\gamma}
\left.\frac{1}{\sqrt 6} A_0 e^{i\delta_0^\pi} -
\left|\frac{V_{td}^* V_{tb}}{V_{ub} V_{ud}^*}\right| e^{i\beta} P
e^{i(\delta_P^\pi  + \tilde\delta_0^\pi)}\right\} \;\; , \label{3}
\end{eqnarray}
\begin{equation}
M_{\bar B_u \to \pi^-\pi^0} = \frac{G_F}{\sqrt 2}|V_{ub} V_{ud}^*|
m_B^2 f_\pi f_+(0) \left\{\frac{\sqrt 3}{2\sqrt 2} e^{-i\gamma}
A_2 e^{i\delta_2^\pi} \right\} \;\; , \label{4}
\end{equation}
where $V_{ik}$ are the elements of CKM matrix, $\gamma$ and $\beta$
are the unitarity triangle angles and we factor out the product $m_B^2
f_\pi f_+(0)$ which appears when the decay amplitudes are calculated
in the factorization approximation. $A_2$ and $A_0$ are the absolute
values of the decay amplitudes into the states with $I=2$ and 0,
generated by operators $O_1$ and $O_2$ (tree amplitudes), while
$P$ is the absolute value of QCD penguin amplitude (generated by
operators $O_3 - O_6$ of effective nonleptonic Hamiltonian which
describes $b$ quark decays into the states without charm and strange
quarks). $\delta_0^\pi$, $\delta_2^\pi$ and $\tilde\delta_0^\pi$
are FSI phases of these three amplitudes, and it is very important
for what follows
that all of them are different. It is easy to understand why
$\delta_0^\pi$ is different from $\delta_2^\pi$: strong
interaction depends on the isospin and is different for $I=0$ and
$I=2$. For example, there are definitely quark-antiquark
resonances with $I=0$, while exotic resonances with $I=2$ should
be made from at least four quarks and their existence is
questionable. The reason why $\delta_0^\pi$ differs from
$\tilde\delta_0^\pi$ is more subtle. Let us consider the intermediate
state made from two charged $\rho$-mesons which contributes to FSI
phases: $B_d \to\rho^+\rho^- \to\pi\pi$. $\rho^+\rho^-$
intermediate state contribution to FSI phases can be large since ${\rm
Br} (B_d\to\rho^+\rho^-)$ is big. Both tree and penguin induced
amplitudes get FSI phases through this chain. Its contribution to
$\delta_0^\pi$ is proportional to $\sqrt{({\rm Br}
B_d\to\rho^+\rho^-)_T/({\rm Br} B_d \to\pi^+\pi^-)_T}
\approx   \sqrt{({\rm Br}
B_d\to\rho^+\rho^-)/({\rm Br} B_d \to\pi^+\pi^-)}
\approx 2.1$,
while that to $\tilde\delta_0^\pi$ is proportional to $\sqrt{({\rm Br}
B_d\to\rho^+\rho^-)_P/({\rm Br} B_d \to\pi^+\pi^-)_P}$.

How can we determine the penguin contributions to the probabilities of
$B_d\to\rho^+\rho^-$ and $B_d\to\pi^+\pi^-$-decays? The most
straightforward way suggested in literature is to extract them
from the probabilities of $B_u\to K^{0*}\rho^+$ and $B_u\to K^0
\pi^+$ decays to which tree amplitudes almost do not contribute
\cite{3,4}\footnote{Contribution of tree amplitudes to these
decays comes from the rescattering $(B_u \to K^+\pi^0)_T$, $K^+
\pi^0 \to K^0 \pi^+$, and taking into account CKM suppression of 
the 
tree amplitudes of $B\to K\pi(K^*\rho)$ decays relative to the penguin
amplitudes we can cautiously estimate
tree contribution as not more than 10\% of penguin one
.}:
\begin{equation}
{\rm Br}(B_d\to\rho^+\rho^-)_P =
\left(\frac{f_\rho}{f_{K^*}}\right)^2
\left[\lambda\sqrt{\eta^2+(1-\rho)^2}\right]^2
\frac{\tau_{B_d}}{\tau_{B_u}} {\rm Br}(K^{0*}\rho^+) \approx 0.34
\cdot 10^{-6} \;\; , \label{5}
\end{equation}
\begin{equation}
{\rm Br}(B_d\to\pi^+\pi^-)_P = \left(\frac{f_\pi}{f_K}\right)^2
\left[\lambda\sqrt{\eta^2+(1-\rho)^2}\right]^2
\frac{\tau_{B_d}}{\tau_{B_u}} {\rm Br}(K^0\pi^+) \approx 0.59
\cdot 10^{-6} \;\; , \label{6}
\end{equation}
where $f_\rho = 209$ MeV and $f_{K^*} =  218$ MeV are the vector meson
decay constants, $\lambda = 0.23$, $\eta = 0.34$ and $\rho = 0.20$
are the CKM matrix parameters in Wolfenstein parametrization \cite{5},
$f_K/f_\pi = 1.2$ and the central values of ${\rm Br}(B_u \to
K^{0*}\rho^+) = (9.2 \pm 1.5) \cdot 10^{-6}$ and ${\rm Br}(B_u \to
K^0\pi^+) = (23.1 \pm 1.0) \cdot 10^{-6}$ \cite{1} were used. The
accuracy of equations (\ref{5}) and (\ref{6}) depends on the accuracy
of $d\leftrightarrow s$ interchange symmetry 
($U$-spin symmetry) of $b\to d(s)$
transition amplitudes described by QCD penguin, however when the ratio 
of
(\ref{5}) to (\ref{6}) is calculated uncertainty factors partially 
cancel out and we obtain rather stable result:
instead of being enchanced as in the case of the tree amplitude
intermediate vector mesons contribution into penguin $B_d
\to\pi^+\pi^-$ amplitude is suppressed,
$(\tilde\delta_0^\pi)_{\rho\rho} \approx 1/2.8
(\delta_0^\pi)_{\rho\rho}$. Taking into account that fraction of
longitudinally polarized vector mesons produced in $B_u\to K^{0*}\rho^+$
decays is about 50\% we get additional suppression of $(\tilde\delta_0^\pi)_{\rho\rho}$ by
factor $\sqrt 2$.

Finally, phase $\delta_P^\pi$ comes from the imaginary part of
the penguin loop with $c$-quark propagating in it \cite{6}. 
In order to calculate $\delta_P^\pi$ let us consider
corresponding quark diagram. The charm penguin contribution is given 
by the
following expression:
\begin{equation}
Pe^{i\delta_P^{\pi}}=
-P_c(k^2)=\frac{1}{3} \ln(\frac{M_W^2}{m_b^2}) +
i\frac{\pi}{3}(1+\frac{2m_c^2}{k^2})\sqrt{1-\frac{4m_c^2}{k^2}}
\;\; , \label{7}
\end{equation}
where $k$ is the sum of momenta of two quarks to which gluon radiated
from penguin 
decays: $k=p_1 + p_2$. One of these quarks forms $\pi$-meson with the
spectator quark, so neglecting spectator quark momentum in the
rest frame of $B$-meson we have $p_1 = (\frac{m_b}{2},
\frac{m_b}{2})$. The second quark forms another $\pi$-meson with
$\bar d$-quark radiated from penguin: $p_2 = x(\frac{m_b}{2},
-\frac{m_b}{2})$ where $0 < x < 1$ is the fraction of $\pi^+$
momentum carried by $u$-quark. Substituting $k^2 = x m_b^2$ into
(\ref{7}) and integrating it with the asymptotic quark distribution function
in $\pi$-meson $\varphi_\pi(x) = x(1-x)$ we obtain the value of
$\delta_P^\pi$ which depends on the ratio $4m_c^2/m_b^2$. In
particular, for $m_b =5.3$ GeV and $m_c = 1.9$ GeV
(which correspond to the masses of physical states) we obtain
$\delta_P^\pi \approx 10^o$, a small positive value. A 
nonperturbative calculation of  $\delta_P^\pi$  
described in Section 3 demonstrates that the sign of
$\delta_P^\pi$ can be negative.

Our next task is to determine the difference of FSI phases
$\delta_0^\pi - \delta_2^\pi$ (the large value of it is responsible
for a relatively small value of $R_\pi$). If we neglect the penguin
contribution, then from (\ref{2}) - (\ref{4}) we get the following
expression:
\begin{equation}
\cos(\delta_0^\pi -\delta_2^\pi) = \frac{\sqrt 3}{4} \frac{{\rm
B}_{+-} - 2 B_{00} + \frac{2}{3} \frac{\tau_0}{\tau_+}
B_{+0}}{\sqrt{\frac{\tau_0}{\tau_+} B_{+0}}\sqrt{B_{+-} + B_{00}
-\frac{2}{3} \frac{\tau_0}{\tau_+} B_{+0}}} \;\; , \label{8}
\end{equation}
where $B_{ik}$'s are the $C$-averaged branching ratios, while
$\tau_0/\tau_+ \equiv \tau(B_d)/\tau(B_u) = 0.92$. Substituting the
central values from Table 1 we get $|\delta_0^\pi - \delta_2^\pi|
= 48^o$.

Penguin contributions to $B_{ik}$ do not interfere with tree ones
because $\alpha = \pi -\beta -\gamma$ is almost equal to $\pi/2$.
Taking $P^2$ terms into account with the help of (\ref{6}) 
(subtracting 0.59 and 0.30 from the first and the second lines of 
Table 1 numbers
describing $B\to\pi\pi$ data correspondingly)
we get:
\begin{equation}
|\delta_0^\pi - \delta_2^\pi| = 37^o \pm 10^o \;\; . \label{9}
\end{equation}
The accuracy of this 11$^o$ decrease of the absolute value of the 
phases
difference is determined by the accuracy of (\ref{6}) and is not 
high. In
recent paper \cite{7} the global fit of $B\to\pi\pi$ and $B\to\pi K$
decay data was made. The tree amplitudes of $B\to\pi\pi$ decays were
designated in \cite{7} by $T$ for $B\to\pi^+ \pi^-$ and by
$C$ for $B\to\pi^0\pi^0$. According to \cite{7} the difference of FSI
phases between $C$ and $T$ equals $\delta_C = -58^o \pm 10^o$,
$|C| = 0.37 \pm 0.05$, $|T| = 0.57 \pm 0.05$ in the units of
$10^4$ eV. The phase shift between the isospin amplitudes is 
determined by
these quantities:
\begin{equation}
\tan(\delta_0 -\delta_2) = \frac{3TC \sin(-\delta_C)}{2T^2 + TC
\cos\delta_C - C^2} \;\; , \label{10}
\end{equation}
and substituting the numbers we obtain:
\begin{equation}
\delta_0 - \delta_2 = 40^o \pm 7^o \;\; , \label{11}
\end{equation}
the result very close to (\ref{9}). However, the same
$d\leftrightarrow s$ interchange symmetry was used in \cite{7}
when relating $B\to\pi\pi$ and $B\to K\pi$ decays. 
Fit \cite{7} was made in the same ``$t$-convention'' which we use
(see the
statement at the end of page 3 of the paper \cite{7}: 
``for simplicity, we will assume
... $P_{tc} = P_{tu}$''), therefore the obtained results can be
directly compared with ours.

Now let us consider $B\to\rho\rho$ decays. According to BABAR and
BELLE results $\rho$ mesons produced in $B$ decays are almost
entirely longitudinally polarized ($f_L(\rho_+\rho_-) = 0.98 \pm
0.03$\cite{80}, $f_L(\rho_+\rho_0) = 0.91 \pm 0.4$ \cite{8},
$f_L(\rho_0\rho_0) = 0.86 \pm 0.12$ \cite{9}). For $B$ decays into
the longitudinally  polarized $\rho$-mesons we can write formulas
analogous to (\ref{2}) - (\ref{4}) and we can find FSI phases
difference with the help of analog of (\ref{8}). Substituting the central values of
branching ratios of $B\to\rho\rho$ decays from Table 1 we obtain:
$|\delta_0^\rho - \delta_2^\rho| = 21^o$. In order to subtract the
penguin contribution with the help of (\ref{5}) we should take
into account that in $B_u\to K^{0*}\rho^+$  decays the fraction 
of the
longitudinally polarized vector mesons equals approximately 50\%
\cite{10}, so we should subtract $0.17 \cdot 10^{-6}$ in case of
decay to $\rho^+\rho^-$ and $0.08 \cdot 10^{-6}$ for decay into
$\rho^0\rho^0$. In this way we obtain:
\begin{equation}
|\delta_0^\rho - \delta_2^\rho| = {20^o}^{+8^o}_{-20^o} \;\; ,
\label{12}
\end{equation}
and the factor 2 difference between (\ref{12}) and (\ref{9}) or
(\ref{11}) is responsible for the different patterns of
$B\to\rho\rho$ and $B\to\pi\pi$ decay probabilities. Let us 
emphasize
that while $|\delta_0^{\rho} - \delta_2^{\rho}|$ being only
one standard deviation from zero
can be very small this is not so for
$|\delta_0^\pi - \delta_2^\pi|$.

\section{Calculation of the FSI phases of $B\to\pi\pi$
and $B\to\rho\rho$ decay amplitudes}

Among three amplitudes of $B\to\pi\pi$ decays (\ref{2})--(\ref{4})
only two are independent. We will calculate FSI phases of
$B\to\pi^+\pi^0$ and $B\to\pi^+\pi^-$ amplitudes and extract from them 
FSI phases of amplitudes with a definite isospin.

Our task is to take into account  the intermediate state
contributions into FSI phases. 
As it was argued in
Introduction we should consider only two particle intermediate
states with positive $G$-parity to which $B$-mesons have relatively
large decay probabilities. Alongside with $\pi\pi$ and $\rho\rho$ 
there is only
one such state: $\pi a_1$. So we will consider $\rho\rho$ 
intermediate state which transforms into $\pi\pi$ by $\pi$ 
exchange in $t$-channel,
$\pi a_1$  intermediate state which transforms into
$\pi\pi$ by $\rho$ exchange in $t$-channel and will take into account
the elastic channel $B\to\pi\pi\to\pi\pi$ as well. 
This approach is analogous to the FSI consideration performed
in paper \cite{111}. However in \cite{111} $2 \rightarrow 2$
scattering amplitudes were considered to be due to
elementary particle exchanges in $t$-channel. For vector particles
exchanges $s$-channel partial wave amplitudes behave as 
$s^{J-1} \sim s^0$ and thus do not decrease with energy (decaying
meson mass). However it is well known that  the correct
behavior is given by Regge theory: $s^{\alpha_i(0)-1}$.
For $\rho$-exchange $\alpha_{\rho}(0) \approx 1/2$ and the
amplitude decrease with energy as $1/\sqrt s $.
This effect
is very spectacular for $B\to DD\to\pi\pi$ chain with $D^*(D^*_2)$
exchange in $t$-channel: $\alpha_{D^*}(0) \approx -1$
and reggeized $D^*$ meson exchange is damped
as $s^{-2} \approx 10^{-3}$ in comparison with elementary $D^*$
exchange (see for example \cite{11}). For $\pi$-exchange, which
gives a dominant contribution to $\rho\rho \to \pi\pi$ transition 
(see below), in the small $t$ region the  
pion is close to mass shell and its reggeization is not important.

We will use Feynman diagram approach to calculate FSI phases from
the triangle diagram with the low mass intermediate states $X$ and
$Y$ (see Figure 1). Integrating over loop momenta $d^4 k$ 
we assume that integrals over masses of intermediate states $X$ and $Y$
decrease
   rapidly with increase of these masses.
 Then choosing $z$ axis in the direction of
momenta of
the produced meson $M_1$ we can transform the integral over
$k_0$ and $k_z$ into the integral over the invariant masses of clusters
of intermediate particles $X$ and $Y$
\begin{equation}
\int dk_0dk_z=\frac{1}{2M_B^2}\int ds_X ds_Y \label{260}
\end{equation}
and deform integration contours in such a way that only low mass
intermediate states contributions are taken into account while the
contribution of heavy states
being small is neglected.
In this way we get:
\begin{equation}
M_{\pi\pi}^I = M_{XY}^{(0)I} (\delta_{\pi X}\delta_{\pi Y} + i
T_{XY \to\pi\pi}^{J=0}) \;\; , \label{26}
\end{equation}
where $M_{XY}^{(0)I}$ are the decay matrix elements without FSI
interactions and $T_{XY\to\pi\pi}^{J=0}$ is the $J=0$ partial wave
amplitude of the process $XY\to\pi\pi \;\; (T^J = (S^J -1)/(2i))$
which originates from the integral over $d^2k_{\perp}$.

\begin{figure}[!htb]
\centering
\epsfig{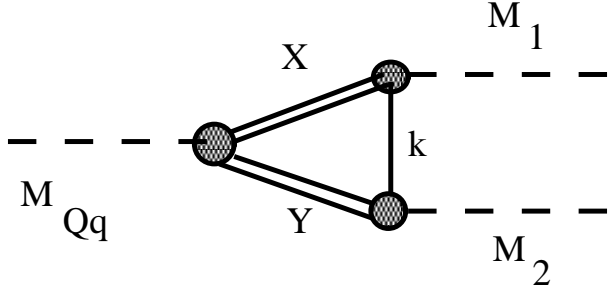}
\caption{\em Diagram which describes FSI in the decay of heavy meson 
$M_{Qq}$ into two light mesons $M_1$ and $M_2$. $X$ and $Y$ are the clusters
of particles with small invariant masses $s_X,s_Y \leq M_Q \Lambda_{QCD}$,
$k$ is $4$-momentum of a virtual particle propagating in $t$-channel. }
\label{WW2Fermi}
\end{figure}

For real $T$  (\ref{26}) coincides with the application of the unitarity
condition for the calculation of the imaginary part of $M$ while for 
the imaginary $T$ the corrections to the real part of $M$ are generated.

Let us calculate the imaginary parts of $B\to\pi\pi$ decay amplitudes
which originate from $B\to\rho\rho\to\pi\pi$ chain with the help
of unitarity condition \footnote{in this section the phases which
originate from CKM matrix elements are omitted.}:
\begin{equation}
{\rm Im} M(B\to\pi\pi) = \int\frac{d\cos\theta}{32\pi}
M(\rho\rho\to\pi\pi) M^* (B\to\rho\rho) \;\; , \label{13}
\end{equation}
where $\theta$ is the angle between $\rho$ and $\pi$ momenta. For
small values of $\theta$ or $t$ $\pi$-exchange in $t$-channel
dominates and the calculation of Feynman diagram for
$\rho\rho\to\pi\pi$ amplitude with the elementary virtual $\pi$-meson
exchange can be trusted, as it was noted above. It was already  stressed that $\rho$-mesons
produced in $B$-decays are almost entirely longitudinally
polarized. That is why we will take into account only longitudinal
polarization for the intermediate $\rho$-mesons and amplitudes of
$B$-decays into $\pi\pi$ and $\rho_L\rho_L$ are simply related
\footnote {relative negative sign of the amplitudes follows from the
expressions for transition formfactors in the factorization approximation,
see for example \cite{A}.}:
\begin{equation}
M_{B^+\to\rho^+\rho^0} = -\sqrt{\frac{18.2}{5.7}}
M_{B^+\to\pi^+\pi^0} \; , \;\; M_{B_d\to\rho^+\rho^-} =
-\sqrt{\frac{23.1}{5.2}} M_{B_d\to\pi^+\pi^-} \;\; . \label{14}
\end{equation}
For the amplitude of $\rho^+\rho^0\to\pi^0\pi^+$ transition
we have:
\begin{equation}
iM(\rho^+\rho^0\to\pi^0\pi^+) = -i \frac{g_{\rho_{\pi\pi}}^2}{(p_1
- k_1)^2 - m_\pi^2} (k_1 \rho^+) (k_2 \rho^0) \;\; , \label{15}
\end{equation}
where $p_1$, $k_1$ and $k_2$ are $\rho^+$, $\pi^0$ and $\pi^+$
momenta. From the width of $\rho$-meson we get
$g^2_{\rho\pi\pi}/16\pi = 2.85$. For the longitudinally polarized
$\rho$-mesons in their center of mass system we have:
\begin{equation}
k_1\rho^+ = k_2 \rho^0 = -\frac{1}{2m_{\rho}}
\left[(t-m_\pi^2)(1+\frac{m_\rho^2}{2E_\rho^2}) +m_\rho^2 \right]
\;\; , \label{16}
\end{equation}
where $t=(p_1 -k_1)^2$. Changing the integration variable in
(\ref{13}) to $t$ with the help of $dt =
\frac{M_B^2}{2}(1-2\frac{m_\rho^2}{M_B^2}) d\cos\theta$ and
introducing formfactor $exp(t/\mu^2)$ with the parameter 
$\mu^2 \sim 1 \; GeV^2$ we obtain:
\begin{eqnarray}
{\rm Im} M_{B\to\pi^+\pi^0} & = & +\sqrt{\frac{18.2}{5.7}}
\left\{\int\limits_{-\infty} ^{-\frac{(m_\rho^2 -
m_\pi^2)^2}{M_B^2}}\frac{g_{\rho\pi\pi}^2 dt}{16\pi M_B^2*
4m_\rho^2} \left[ (t-m_\pi^2)(1+\frac{2m^2_{\rho}}{M_B^2})^2 +
\right.\right. \nonumber
\\
& + & \left.\left. 2m_\rho^2 (1+\frac{2m^2_\rho}{M_B^2}) +
\frac{m_\rho^4}{t-m_\pi^2}\right] exp(t/\mu^2)\right\}
M_{B\to\pi^+\pi^0} \;\; . \label{17}
\end{eqnarray}
For $\mu^2 = 2m_\rho^2$ the contributions of the first two terms in
square brackets cancel, while the third term gives:
\begin{equation}
{\rm Im} M_{B\to\pi^+\pi^0} = -\sqrt{\frac{18.2}{5.7}}
\frac{g_{\rho\pi\pi}^2}{16\pi} \frac{m_\rho ^2}{4 M_B^2} 3.1
M_{B\to\pi^+\pi^0} \;\; , \label{18}
\end{equation}
and from (\ref{4}) we get:
\begin{equation}
\delta_2^\pi(\rho\rho) = -4.9^o \;\; . \label{19}
\end{equation}
Let us note that in the limit $M_B \to\infty$ 
the ratio $Br(B_d\to\rho\rho)/Br(B_d\to\pi\pi)$ grows as $M^2_B$,
that is why  FSI phase
$\delta_2^\pi(\rho\rho)$  (and  $\delta_0^\pi(\rho\rho)$) 
diminishes as $1/M_B$.

The analogous consideration of $\rho^+\rho^-$ intermediate state leads
to the positive FSI phase of $B_d\to\pi^+\pi^-$ amplitude which is
enhanced relatively to $\delta_2^{\pi}(\rho\rho)$ according to
(\ref{14}):
\begin{equation}
\delta_{+-}^\pi(\rho\rho) = +5.7^o \;\; , \label{20}
\end{equation}
and for FSI phase of the amplitude with isospin zero in the linear
approximation we get:
\begin{equation}
\delta_0^\pi(\rho\rho) = \delta_{+-}^\pi(\rho\rho)
+\frac{A_2}{\sqrt{2} A_0} \left[ \delta_{+-}^\pi(\rho\rho) -
\delta_2^\pi(\rho\rho)\right] \;\; . \label{21}
\end{equation}
We are able to extract the ratio $A_2/A_0$ from that of $C$-averaged 
${\rm Br}(B_d\to\pi^+\pi^-)$, ${\rm Br} (B_d\to\pi^0\pi^0)$
and ${\rm Br}(B_u \to\pi^+\pi^0)$, subtracting penguin
contribution as we did deriving  (\ref{9}):
\begin{equation}
\left(\frac{A_0}{A_2}\right)^2 = \frac{\tilde B_{+-} + \tilde
B_{00}}{\frac{2}{3} B_{+0} \frac{\tau_0}{\tau_+}} -1 \;\; ,
\label{22}
\end{equation}
\begin{equation}
\frac{A_0}{A_2} = 0.80 \pm 0.09 \;\; , \label{23}
\end{equation}
and, finally:
\begin{equation}
\delta_0^\pi(\rho\rho) = 15^o \; , \;\; \delta_0^\pi(\rho\rho) -
\delta_2^\pi(\rho\rho) =
20^o \;\; . \label{24}
\end{equation}

In this way we see that $B\to\rho\rho\to\pi\pi$ chain generates
half of the experimentally observed FSI phase difference of 
$B\to\pi\pi$ tree amplitudes.

It is remarkable that FSI phases generated by
$B\to\pi\pi\to\rho\rho$ chain are damped by ${\rm
Br}(B\to\rho^+\rho^-, \rho^+\rho^0)/{\rm Br}(B\to\pi^+\pi,
\pi^+\pi^0)$ ratios and are a few degrees: $$ \delta_2^{\rho}(\pi\pi)=
 \frac{5.7}{18.2} * \delta_2^\pi(\rho\rho)=-1.4^o \; ,
\;\; \delta_{+-}^\rho(\pi\pi) = \frac{5.2}{23.1} *
\delta_{+-}^\pi(\rho\rho) = 1.2^o \; , $$
\begin{equation}
(A_0/A_2)_{\rho\rho}  =  1.1 \; , \;\; \delta_0^\rho(\pi\pi)=2.9^o
\; , \;\; \delta_0^\rho(\pi\pi) - \delta_2^\rho(\pi\pi) \approx
4^o \; . \label{25}
\end{equation}

Next we will take into account $\pi\pi$ intermediate state.
From Regge analysis of
$\pi\pi$ elastic scattering we know that good description
of the experimental data is achieved when the exchanges of pomeron, 
$\rho$ and $f$
trajectories in $t$-channel are taken into account \cite{12}.   
Pomeron
exchange dominates in elastic $\pi\pi\to\pi\pi$ scattering at high
energies. For $\alpha_P(0)=1$ the corresponding amplitude $T$ is
purely imaginary and the phases of matrix elements 
do not change \cite{2}.
However taking into account that pomeron is "supercritical",
$\alpha_P(0)\approx 1.1$, we obtain the phase of the amplitude 
generated by pomeron exchange
 \footnote{The amplitude of $2\to 2$
process due to supercritical pomeron exchange is $T\sim (s/s_0)^{
\alpha_P(t)}(1+exp(-i\pi\alpha_P(t)))/(-\sin(\pi\alpha_P(t)))=(s/s_0)^
{(1+\Delta)}(i+\Delta\pi/2)$, where in the last expression $t=0$ was substituted and $\alpha_P(0)=1+\Delta$ was used ($\Delta\approx0.1$).}
which cancels the
phases generated by $\rho$ and $f$ exchanges for $I=2$. For $I=0$
the sum of  $\rho$ and $f$ exchanges produces the purely imaginary amplitude
$T$ and the phase of the amplitude $M$ is due to
pomeron "supercriticallity":

\begin{equation}
\delta_0^\pi(\pi\pi) = 5.0^o \; , \;\delta_2^\pi(\pi\pi) = 0^o\;\; 
\label{261}.
\end{equation}

In paper \cite{2} the pomeron exchange amplitude was considered as purely
imaginary. As a result though important for branching ratios phase 
difference $\delta_0^\pi(\pi\pi) - \delta_2^\pi(\pi\pi)$ was the same 
(pomeron contribution being universal cancels in the difference of phases)
it came mainly from  $\delta_2^\pi(\pi\pi)$ negative value.
In this way result for the absolute value of direct CP-asymmetry 
$C_{\pi^+\pi^-}$ was underestimated, see below. 

Finally $\pi a_1$ intermediate state should be accounted for. 
Large branching ratio of $B_d\to\pi^{\pm} a_1^{\mp}$-decay
( ${\rm Br}(B_d\to\pi^{\pm} a_1^{\mp})=(40 \pm 4)*10^{-6}$) is
partially compensated by small $\rho\pi a_1$ coupling constant
(it is $1/3$ of $\rho\pi \pi$ one).
As a result the contributions of  $\pi a_1$ intermediate state (which
transforms into $\pi\pi$ by $\rho$-trajectory exchange in $t$-channel)
to FSI phases equal approximately that part of $\pi\pi$ intermediate
state contributions which is due to $\rho$-trajectory exchange.
Assuming that the sign of the $\pi a_1$ intermediate state contribution into
phases is the same as that of elastic channel we obtain:  

\begin{equation}
\delta_0^\pi(\pi a_1) = 4^o \; , \;\delta_2^\pi(\pi a_1) = -2^o\;\; 
\label{262}.
\end{equation}

Summing the imaginary parts of the amplitudes which follow from
(\ref{19}), (\ref{24}), (\ref{261}) and (\ref{262})
we finally obtain:
\begin{equation}
\delta_0^\pi = 23^o \; , \;\;\delta_2^\pi = -7^o\;\;,\;\;
\delta_0^\pi -
\delta_2^\pi =
30^o \;\; , \label{263}
\end{equation}
and the accuracy of these numbers is not high, at the level of $50\%$.

The analogous consideration of the real parts of the loop corrections
to $B\to\pi\pi$ decay amplitudes
leads to the diminishing of the (real) tree amplitudes by $\approx 30\%$,
and
we can explain the
experimentally observed value $\delta_0^\pi -
\delta_2^\pi\approx 40^o$ in our model while for $\rho\rho$ final
state the analogous difference is about three times smaller, $\delta_0^\rho - \delta_2^\rho \approx
15^o $.

Let us estimate the phase of the penguin amplitude $\delta^{\pi}_P$
considering charmed mesons intermediate states: $B\to \bar D D,
\bar{D^*} D, \bar D D^*, \bar{D^*} D^* \to \pi\pi$
\footnote{These amplitudes are considered as penguin  
due to the proper combination of CKM matrix
elements.}. In Regge model
all these amplitudes are described at high energies by exchanges of
$D^*(D^*_2)$-trajectories. An intercept of these exchange-degenerate
trajectories can be obtained using the method of \cite{k} or from masses of $D^*(2007) \; $--$\; 1^-$ and $D^*_2(2460) \;$--$\; 2^+$ resonances,
assuming  linearity of these Regge-trajectories. Both methodes give
$\alpha_{D^*}(0) = -0.8 \div -1$ and the slope $ \alpha_{D^*}'
\approx 0.5 GeV^{-2}$.

The amplitude of $D^+D^-\to\pi^+\pi^-$ reaction in the Regge model 
proposed in papers \cite{v,b} can be written in the following form:
\begin{equation}
T_{D\bar D \to\pi\pi}(s,t)=-\frac{g^2_0}{2}e^{-i\pi\alpha(t)}
\Gamma(1-\alpha_{D^*}(t))(s/s_{cd})^{\alpha_{D^*}(t)} \;\; ,
\label{1259}
\end{equation}
where $\Gamma (x)$ is the gamma function.

The $t$-dependence of Regge-residues is chosen in accord with the dual 
models and is tested for light (u,d,s) quarks \cite{v}. According to
\cite{b} $s_{cd} \approx 2.2 GeV^2$. 

Note that the sign of the amplitude is fixed by the unitarity in the 
$t$-channel (close to the $D^*$-resonance). The constant $g^2_0$ is
determined by the width of the $D^* \to D \pi$ decay: $g^2_0/(16\pi)=6.6$.
Using (\ref{26}), analog of (\ref{13}), (\ref{1259}) and the branching
ratio $Br (B\to D \bar D)\approx 2\cdot 10^{-4}$ \cite{p} we obtain
the imaginary part of $P$ and comparing it with the contribution of
$P$ in $B\to \pi^+\pi^-$ decay probability (\ref{6}) we get 
$\delta^{\pi}_P \approx -3.5^o$\footnote{In integration over
$\cos\theta$ the region $\theta \ll 1$ dominates.
In this region representation (\ref{1259}) is valid.}. A smallness of the phase is due to the low
intercept of $D^*$-trajectory. The sign of $\delta_P$ is negative - opposite
to the positive sign which was obtained in perturbation theory (\ref{7}).

Since $D\bar D$-decay channel constitutes only $\approx 10\%$ of all two-body
charm-anticharm decays of $B_d$-meson \cite{p} taking these channels into
account we can easily get 
\begin{equation}
\delta_P \sim -10^o \;\; ,
\label{1260}
\end{equation}
which may be very important for the interpretation of the experimental data
on direct CP asymmetry $C_{+-}$ discussed in the next section.

\section{CP asymmetries of $B_d(\bar B_d)\to\pi\pi$ decays}

The CP asymmetries are given by : 
\begin{equation}
 C_{\pi\pi} \equiv
\frac{1-|\lambda_{\pi\pi}|^2}{1+|\lambda_{\pi\pi}|^2} \; , \;\;
S_{\pi\pi} \equiv \frac{2{\rm Im}(\lambda_{\pi\pi})}{1+
|\lambda_{\pi\pi}|^2} \; , \;\; \lambda_{\pi\pi} \equiv
e^{-2i\beta} \frac{M_{\bar B \to \pi\pi}}{M_{B\to\pi\pi}} \;\; ,
\label{259} 
\end{equation}
where $\pi\pi$ is $\pi^+\pi^-$ or $\pi^0 \pi^0$.

From (\ref{2}) for direct CP asymmetry in $B_d(\bar B_d) \to \pi^+\pi^-$
decays we readily obtain:
\begin{eqnarray}
C_{+-} & = & -\frac{\tilde P}{\sqrt 3} \sin\alpha [\sqrt 2 A_0
\sin(\delta_0 - \tilde\delta_0 - \delta_P) + A_2 \sin(\delta_2 -
\tilde\delta_0 -\delta_P)]/ \nonumber \\ & / &[\frac{A_0^2}{6} +
\frac{A_2^2}{12} + \frac{A_0 A_2}{3\sqrt 2} \cos(\delta_0 -
\delta_2) -
\sqrt{\frac{2}{3}} A_0 \tilde P \cos\alpha \cos(\delta_0 -
\tilde\delta_0 - \delta_P) - \nonumber \\ & - & \frac{A_2 \tilde
P}{\sqrt 3} \cos\alpha \cos(\delta_2 - \tilde\delta_0 - \delta_P)
+ \tilde P^2] \;\; , \label{264}
\end{eqnarray}
where
\begin{equation}
\tilde P \equiv \left|\frac{V_{td}^* V_{tb}}{V_{ub}
V_{ud}^*}\right| P \;\; . \label{265}
\end{equation}

In order to make a numerical estimate we should know the ratios 
$A_0/A_2$ and $P/A_2$. The first one is given by (\ref{23}) while the
second one can be extracted from the ratio ${\rm Br}(B_u\to K^0\pi^+)/
{\rm Br}(B_u\to \pi^0\pi^+)$ assuming $d \leftrightarrow s$ invariance of
the strong interactions:

\begin{equation}
\frac{{\rm Br}(B_u\to K^0\pi^+)}{{\rm Br}(B_u\to \pi^0\pi^+)} = 
\frac{f^2_K P^2|V^*_{ts}
V_{tb}|^2}{f^2_{\pi} \frac{3}{8}A_2^2|V^*_{ud} V_{ub}|^2} \;\; , \label{266}
\end{equation}

\begin{equation}
\frac{P}{A_2}=0.092(0.009) \;\; .\label{267}
\end{equation}

The numerical values of $A_0$ and $A_2$ are given with good accuracy
by factorization calculation, while $P$ appears to be 2.5 times
larger than factorization result \cite{2}. In view of this the validity 
of factor $f_K$ in (\ref{266}) which originates from
factorization calculation of the penguin amplitude is questinable. 
If factorization of the penguin amplitudes is not assumed then the
ratio $f_{K}/f_{\pi}$ in (\ref{266}) should be replaced by
unity. In this way we get $20\%$ larger value of $P/A_2$ in
(\ref{267}) and we will take this value of uncertainty as an estimate
of the theoretical accuracy of the determination of $P$:

\begin{equation}
\frac{\tilde P}{A_2}=0.21(0.04) \;\; ,\label{268}
\end{equation}
Taking into account that unitarity 
triangle angle $\alpha \approx 90^o $ and angles 
$\tilde \delta_0$ and $\delta_P$ are of the order of few
degrees from (\ref{264}) we obtain:
\begin{eqnarray}
C_{+-} &\approx & -0.28[1.1\sin(\delta_0 - \tilde\delta_0 - \delta_P)+
\sin(\delta_2 - \tilde\delta_0 - \delta_P)] \approx \nonumber \\
&\approx & -0.56 \sin(
(\delta_0+\delta_2)/2 - \tilde\delta_0 - \delta_P) \;\; .\label{269}
\end{eqnarray}

In order to determine the lower bound on the value of $C_{+-}$ let us suppose that
$\delta_0=37^o, \delta_2=0^o$ (we keep the difference
$\delta_0 - \delta_2=37^o$, as it follows from the data on 
$ B \to \pi \pi $ decay probabilities (\ref{9})), and neglect
small values of $\tilde\delta_0$ and $ \delta_P$:
\begin{equation}
C_{+-} > -0.18 \;\; .\label{270}
\end{equation}

Concerning experimental number it could well happen that finally it
will be considerably below our bound. In this case the result
of nonperturbative calculation of penguin phase will be
confirmed. Substituting in (\ref{269}) $\delta_0=30^o, \delta_2=-7^o$   and $ \delta_P$ from (\ref{1260}) we obtain the following central
value:
\begin{equation}
C_{+-} = -0.21 \;\; .\label{2700}
\end{equation}

It is instructive to compare the obtained numbers with the value of $C_{+-}$
which follows from the asymmetry $A_{CP}(K^+ \pi^-)$ if $d \leftrightarrow s$
symmetry is supposed \cite{13a}:
\begin{eqnarray}
C_{+-} & = & \left(\frac{f_{\pi}}{f_K}\right)^2 A_{CP}(K^+ \pi^-)
\frac{\Gamma(B\to K^+ \pi^-)}{\Gamma(B\to \pi^+ \pi^-)} \frac{
\sin(\beta + \gamma)}{\sin(\gamma)}\left|\frac{V_{td}}{V_{ts}\lambda}
\right| = \nonumber \\
& = & 1.2^{(-2)}(-0.093 \pm 0.015)\frac{19.8}{5.2} \frac{\sin82^o}{\sin60^o}
0.87=-0.24 \pm 0.04 \;\; .\label{271}
\end{eqnarray}

Let us note that one factor $f_{\pi}/f_K$ in the last equation
appears from the matrix element of the tree operator, the second one -
from the matrix element of the penguin operator. If 
because of nonfactorization of penguin amplitudes
we will omit the factor which appears from the penguin \cite{4},
then the
numbers in the right-hand sides of  (\ref
{270}, \ref{2700}) and (\ref{271})  will become
$20\%$ smaller.

The experimental results obtained by Belle \cite{14} and BABAR \cite{15}
are contradictory
\begin{equation}
C_{+-}^{Belle}=-0.55(0.09) \;\; , C_{+-}^{BABAR}=-0.21(0.09),     \label{272}
\end{equation}
Belle number being far below (\ref{270}) and (\ref{2700}).

For direct CP asymmetry in $B_d(\bar B_d)\to \pi^0\pi^0$ decay
from (\ref{3}) we readily obtain:

\begin{eqnarray}
C_{00} & = & -\sqrt{\frac{2}{3}} \tilde P \sin\alpha [A_0
\sin(\delta_0 - \tilde\delta_0 -\delta_P) - \sqrt 2 A_2
\sin(\delta_2 - \tilde\delta_0 -\delta_P)] / \nonumber \\
& / & [\frac{A_0^2}{6} + \frac{A_2^2}{3} - \frac{\sqrt 2}{3} A_0 A_2
\cos(\delta_0 - \delta_2) - \sqrt{\frac{2}{3}} A_0 \tilde P \cos\alpha
\cos(\delta_0 - \tilde\delta_0 - \delta_P) + \nonumber \\ & + &
\frac{2}{\sqrt 3} A_2 \tilde P \cos\alpha \cos(\delta_2 -
\tilde\delta_0 -\delta_P) + \tilde P^2] \;\; , \label{273}
\end{eqnarray}
\begin{equation}
C_{00} \approx  -1.06[0.8\sin(\delta_0 - \tilde\delta_0 - \delta_P)-
1.4\sin(\delta_2 - \tilde\delta_0 - \delta_P)] \approx -0.6
\;\; , \label{274}
\end{equation}
considerably smaller than $C_{+-}$. This unusually large direct CPV
(measured by $|C_{00}|$) is intriguing task for future measurements
since the present experimental error is too big:

\begin{equation}
C_{00}^{exper}=-0.36(0.32) \;\; .     \label{275}
\end{equation}

Belle and BABAR agree now on the value of another  CPV asymmetry measured in $B_d(\bar B_d)\to \pi^+\pi^-$
decays: $S_{+-}^{exper}= -0.62 \pm 0.09$ \cite{14,15}. From this measurement the value of 
unitarity triangle angle $\alpha$ can be extracted. Neglecting the penguin contribution we get:
\begin{equation}
\sin 2\alpha^{\rm T} = S_{+-} \;\; , \label{276}
\end{equation}
\begin{equation}
 \alpha^{\rm T}=109^o \pm 3^o\;\; . \label{277}
\end{equation}
Penguin shifts the value of $\alpha$. The accurate formula
looks like:
\begin{eqnarray}
S_{+-} & = &[\sin 2\alpha(\frac{A^2_0}{6}
+ \frac{A^2_2}{12} + \frac{A_0 A_2}{3\sqrt 2}\cos(\delta_0 -
\delta_2)) - \nonumber \\ & - &\frac{A_2 \tilde
P}{\sqrt 3} \sin\alpha \cos(\delta_2 - \tilde\delta_0 - \delta_P)
-
\sqrt{\frac{2}{3}} A_0 \tilde P \sin\alpha \cos(\delta_0 -
\tilde\delta_0 - \delta_P)]
/ \nonumber \\ & / & [\frac{A_0^2}{6} +
\frac{A_2^2}{12} + \frac{A_0 A_2}{3\sqrt 2} \cos(\delta_0 -
\delta_2) -
\sqrt{\frac{2}{3}} A_0 \tilde P \cos\alpha \cos(\delta_0 -
\tilde\delta_0 - \delta_P) - \nonumber \\ & - & \frac{A_2 \tilde
P}{\sqrt 3} \cos\alpha \cos(\delta_2 - \tilde\delta_0 - \delta_P)
+ \tilde P^2] \;\; , \label{278}
\end{eqnarray}
and since all the phase shifts are not big the values of cosines 
in 
(\ref{278})
are rather stable relative to their variations. For 
numerical estimates we take $\delta_0=30^o$,
 $\delta_2=-7^o$ and neglect $\tilde\delta_0$ and $\delta_P$.
In this way we get:
 \begin{equation}
 (\alpha)_{\pi\pi} = 88^o \pm 4^0(exper) \pm 5^0 (theor)  \;\; , 
  \label{301}
\end{equation}
where the first error comes from uncertainty in $S_{+-}^{exper}$
while the second one comes from that in the value of penguin
 amplitude, (\ref{268}).
Relatively large theoretical uncertainty in the value of 
$\tilde P$ does not prevent to determine $\alpha$ with good
precision.

The relative smallness of penguin contribution to $B\to\rho\rho$
decay amplitudes allow us to determine  $\alpha$ with better 
theoretical accuracy 
from the experimental measurement of $(S_{+-})_{\rho\rho}$ just as
it was done in \cite{140}. With the help 
of (\ref{5}) we obtain:
\begin{equation}
(\frac{\tilde P}{A_2})_{\rho\rho}=0.060(0.012) \;\; ,\label{700}
\end{equation}
where the same $20\%$ uncertainty in extracting penguin amplitude is
supposed. Using the ratio $(A_0/A_2)_{\rho\rho}$ determined in (\ref{25})
from the (\ref{278}) neglecting strong phases (which are much smaller than
in the case of $B\to \pi\pi$ decays) and taking into account the recent
experimental result
$(S_{+-}^{exper})_{\rho\rho} = -0.06 \pm 0.18$ \cite{1} we obtain:

\begin{equation}
 (\alpha)_{\rho\rho} = 87^o \pm 5^0(exper) \pm 1^0 (theor)  \;\; .
   \label{701}
\end{equation}

Let us point out that considerably larger theoretical error quoted in
\cite{3} follows from the larger theoretical uncertainty in the value 
of penguin
amplitude assumed in that paper.

Our results for $\alpha$ should be compared with the numbers 
which follow from the global fit
of unitarity triangle \cite{5,5a}:
\begin{equation}
 \alpha^{CKM fitter} =(99.0^{+4.0}_{-9.4})^o  \;\; ,  
\alpha^{UTfit} =(93 \pm 4)^o \;\; . \label{302}
\end{equation}
   
We conclude this section with the prediction for the value
of CPV
asymmetry $S_{00}$:
\begin{eqnarray}
S_{00} & = &[\sin 2\alpha(\frac{A^2_0}{6}
+ \frac{A^2_2}{3} - \frac{\sqrt 2 A_0 A_2}{3}\cos(\delta_0 -
\delta_2)) + \nonumber \\ & + &\frac{2 A_2 \tilde
P}{\sqrt 3} \sin\alpha \cos(\delta_2 - \tilde\delta_0 - \delta_P)
-
\sqrt{\frac{2}{3}} A_0 \tilde P \sin\alpha \cos(\delta_0 -
\tilde\delta_0 - \delta_P)]
/ \nonumber \\ & / & [\frac{A_0^2}{6} +
\frac{A_2^2}{3} - \frac{\sqrt 2 A_0 A_2}{3} \cos(\delta_0 -
\delta_2) -
\sqrt{\frac{2}{3}} A_0 \tilde P \cos\alpha \cos(\delta_0 -
\tilde\delta_0 -\delta_P) + \nonumber \\ & + & \frac{2 A_2 \tilde
P}{\sqrt 3} \cos\alpha \cos(\delta_2 - \tilde\delta_0 - \delta_P)
+ \tilde P^2] = 0.70 \pm 0.15 \;\; , \label{305}
\end{eqnarray}
a large asymmetry with the sign opposite to that of $S_{+-}$.

\section{Conclusions}

FSI appeared to be very important in $B\to \pi\pi$ decays.
 
The description of these interactions 
presented in the paper allows to explain the experimentally observed difference of the
ratios of decay probabilities to the 
neutral and charged modes in $B\to \pi\pi$ and $B\to \rho\rho$
decays.

Rather large absolute value of direct CP asymmetry
$C_{+-}$ (if confirmed experimentally) will be a manifestation of the negative sign of penguin FSI phase in accord with nonperturbative calculation
and opposite  to perturbative result.

We are grateful to L.V.Akopyan for checking formulas,
Jose Ocariz for recommendation to include the result for angle
$\alpha$ which follows from CP asymmetry $(S_{+-})_{\rho\rho}$
and M.B.Voloshin for useful discussion.

This work was supported by Russian Agency of Atomic Energy;

A.K. was partly supported by grants RFBR 06-02-17012,
RFBR 06-02-72041-MNTI,
INTAS 05-103-7515 and state contract  02.445.11.7424;

M.V. was partly supported by grants RFBR 05-02-17203 and \newline
NSh-5603.2006.2.


\begin{thebibliography}{150}
\bibitem{1}
HFAG, http://www.slac.stanford-edu/xorg/hfag.
\bibitem{7}
C.-W. Chiang, Y.-F. Zhou, JHEP {\bf 0612} (2006) 027.
\bibitem{2}
A.B. Kaidalov, M.I. Vysotsky, hep-ph/0603013, accepted in Yad. Fiz.
\bibitem{3}
M. Beneke, M. Gronau, J. Rohrer, M. Spranger, Phys.Lett. {\bf B638}
(2006) 68.
\bibitem{4}
M. Gronau, J.L. Rosner, Phys. Lett. {\bf B595} (2004) 339.
\bibitem{5}
CKM fitter, http://ckmfitter.in2p3.fr.
\bibitem{5a}
UTfit, http://utfit.roma1.infn.it.
\bibitem{6}
M. Bander, D. Silverman and A. Soni, Phys. Rev. Lett. {\bf 43},
(1979) 242; \\ G.M. G\'{e}rard and W.-S. Hou, Phys. Rev. {\bf D43},
(1991) 2909.
\bibitem{80}
B. Aubert et al., BABAR Collaboration, hep-ex/0607098 (2006).
\bibitem{8}
B. Aubert et al., BABAR Collaboration, Phys. Rev. Lett. 
{\bf 97} (2006) 261801.
\bibitem{9}
B. Aubert et al., BABAR Collaboration, hep-ex/0607097 (2006).
\bibitem{10}
J. Zhong et al. Belle Collaboration, Phys. Rev. Lett. {\bf 95}
(2005) 141801; \\ B. Aubert et al., BABAR Collaboration,
Phys. Rev. Lett. {\bf 97} (2006) 201801.
\bibitem{111}
H-Y. Cheng, C-K. Chua and A.Soni, Phys. Rev. {\bf D71} (2005)
014030.
\bibitem{11}
A.Deandrea et al., Int. J. Mod. Phys. (2006) 4425.
\bibitem{A}
R.Aleksan et al., Phys. Lett. {\bf B356} (1995) 95.
\bibitem{12} 
K.G.Boreskov, A.A.Grigoryan, A.B.Kaidalov, I.I.Levintov,
Yad. Fiz. {\bf 27},  (1978) 813.
\bibitem{k}
A.B.Kaidalov, Zeit. fur Phys. {\bf C12}, (1982) 63.
\bibitem{v}
P.E.Volkovitsky, A.B.Kaidalov, Sov.J.Nucl.Phys. {\bf 35},
(1982) 909.
\bibitem{b}
K.G.Boreskov, A.B.Kaidalov, Sov.J.Nucl.Phys. {\bf 37},
(1983) 100.
\bibitem{p}
Review of Particle Physics, W.-M. Yao et al., Journal of Physics
{\bf G 33}, (2006) 1.
\bibitem{13a}
R.Fleischer, Phys. Lett. {\bf B459}, (1999) 306.
\bibitem{14}
H.Ishino, Belle, talk at ICHEP06, Moscow (2006).
\bibitem{15}
B.Aubert et al, BABAR Collaboration, hep-ex/0703016 (2007). 
\bibitem{140}
M.I.Vysotsky, Yad. Fiz. {\bf 69}, (2006) 703.

\end{thebibliography}
\end{document}